\documentclass[12pt]{amsart}
\usepackage{temp}

\begin{document}

\title{Spatiotemporal risk prediction for infectious disease spread and mortality}
\author{Catherine Li}
\author{Daniel Lazarev}
\address{Department of Mathematics, Massachusetts Institute of Technology, Cambridge, MA 02139}
\email{cateli@mit.edu}
\email{dlazarev@mit.edu}
\date{}

\maketitle

\begin{abstract}
With the outbreak of the COVID-19 pandemic, various studies have focused on predicting the trajectory and risk factors of the virus and its variants. Building on previous work that addressed this problem using genetic and epidemiological data, we introduce a method, Geo Score, that also incorporates geographic, socioeconomic, and demographic data to estimate infection and mortality risk by region and time. We employ gradient descent to find the optimal weights of the factors' significance in determining risk. Such spatiotemporal risk prediction is important for informed public health decision-making so that individuals are aware of the risks of travel during an epidemic or pandemic, and, perhaps more importantly, so that policymakers know how to triage limited resources during a crisis.
We apply our method to New York City COVID-19 data from 2020, predicting ZIP code-level COVID-19 risk for 2021. 
\end{abstract}

\section{Introduction}\label{sec:intro}
The field of epidemiology examines the incidence, spread, and control of health-related events such as disease. The roots of the field extend to the time of Hippocrates, almost 2500 years ago. In the late 20th century, epidemiology was harnessed to fully eradicate smallpox worldwide \cite{hen72}. The emergence of new infectious diseases in the past few decades, such as the Ebola virus, Zika virus, HIV/AIDS, and most recently, the outbreak of SARS-CoV-2, have prompted more epidemiologic study. Uncovering the source, nature, and significant risk factors of disease spread through epidemiology is thus important to understanding and preparing for future epidemics.

Many approaches can be taken to model the spread of disease mathematically, including least-squares regression, maximum entropy, and Bayesian inference \cite{IveBoz21,WuSunLin22,AnsSorGhoWhi22,HuGonGub17}. Here we introduce the most basic model for infectious disease spread: the SIR compartmental model. This model divides the population into three compartments: $S,$ the number of susceptible individuals; $I,$ the number of infected individuals; and $R,$ the number of recovered individuals, all functions of the time $t$ \cite{She20}. Letting the size of the population be $N$ and ignoring births and deaths, we have $$S+I+R=N.$$ We can also relate $S(t), I(t),$ and $R(t)$ to each other with several differential equations. 

First, we have the equation governing the infection of susceptible individuals: $$\dv{S}{t}=-\frac{\beta SI}{N},$$ where $\beta$ is the rate at which an infected individual infects susceptible individuals (the probability of both meeting someone in $S$ and infecting them). 

Now let $\gamma$ be the average rate at which infected individuals recover (people from $I$ move into $R$). Then $$\dv{R}{t}=\gamma I.$$

Finally, the rate of change in $I$ is obtained by subtracting the rate of decrease in $I$ from its rate of increase, which gives us $$\dv{I}{t}=\frac{\beta SI}{N}-\gamma I.$$

Additionally, $R_0,$ the basic reproduction number, is defined by the ratio $\beta/\gamma.$ This model provides a framework through which disease spread can be studied and understood.

In Section~\ref{sec:prev}, we cover the background of disease spread modeling techniques, focusing on Maher et al.'s evaluation of SARS-CoV-2 mutations based on epidemiological, evolutionary, immune, transmissibility, and language model variables \cite{Mah22}. Their creation of the Epi Score inspires our construction of the Geo Score, explained in Section~\ref{sec:theory}. In Section~\ref{sec:results}, we compare and analyze our Geo Score results using 2020 data against the ground truth of COVID cases and deaths from 2021, across the ZIP codes of New York City. Next, in Section~\ref{sec:weights}, we introduce the idea of tunable weights to minimize error. In Section~\ref{sec:graddesc}, we overview the gradient descent method and implementation, as well as the algorithm's results as compared to a linear regression model. Finally, we reflect on the virtues and limitations of the gradient descent model for tuning scores, suggesting new directions for future research in Section~\ref{sec:discussion}.

\section{Previous Works}\label{sec:prev}
Numerous methods have been used to model the spread of infectious diseases, determine its relevant factors, and estimate properties such as the reproductive number, as several studies did for Lumpy Skin Disease and COVID-19 \cite{AlkVan16,Tao20}. One such method is maximum entropy (MaxEnt), which yields the least biased probability distribution given a set of constraints \cite{Dud07}. Researchers often combine MaxEnt with variations of the SIR model, reviewed in Section \ref{sec:intro}, in order to analyze spatiotemporal disease prevalence \cite{Ges06,LucDeiGri20,NieLiuZha21}.

This idea is explored by Ansari et al. in their paper on the source and path of COVID-19 spread \cite{AnsSorGhoWhi22}. The study directly takes an input of the prior belief of parameters in disease spread, then constrains the model on real-world observations. The researchers used MaxEnt to reweight the possible disease trajectories, yielding an average that fit the observed data up to reasonable error. They coupled maximum entropy with the epidemic model Susceptible-Exposed-Asymptomatic-Infected-Removed (SEAIR). This allowed them to determine the source and path of the outbreak.

Leung describes the mechanics of disease transmission, listing different modes of transmission, as well as viral, environmental, and host determinants of transmission: 
these include host receptor binding affinity, temperature, humidity, vaccination, and social contact patterns \cite{Leu21}. 

Studies have incorporated the above factors into disease spread models. For example, social contact patterns and human mobility patterns (on multiple scales) have significant implications for the spread of infectious diseases and can be studied through social network connections and residence data \cite{Bal09,BelGeiBro11,CaoCheZhe21,DorRamSwa21,KanGaoLia20,Kuc22,Smi21}.

Other studies have determined the ecological niche of viruses in terms of geographical factors such as temperature, vegetation, precipitation, and urbanization to best predict the paths that infectious diseases will take \cite{AlkVan16,WanLiuBer21}.

Another factor of viral disease spread is the structure of the virus in relation to the host's receptor cells. For the SARS-CoV-2 virus, the Spike protein's interaction with the human ACE2 receptor has been studied to determine infectivity \cite{Cha22,Liu21,McC22}.

Finally, viral mutations and escape, which allows new variants of viruses to circumvent immunity from vaccines, are an important area of study in disease spread. In their 2022 study, Maher et al. combine three epidemiological factors into a score that accurately forecasts the spread of SARS-CoV-2 mutations months in advance, furthering the understanding of mutational drivers of future variants of concern, as designated by the CDC \cite{Mah22}. 

The researchers studied a set of mutations—found in over 4 million sequences—with significant ``spreading,'' defined by the magnitude of frequency fold changes in multiple countries, before and after a chosen date. Out of evolution, immune, epidemiology, transmissibility, and language model factors \cite{Hie21}, the epidemiological feature group performed best in predicting the spread of these amino acid mutations, especially in the receptor binding domain (RBD) of the Spike protein. The four variables describing epidemiology were as follows: 
\begin{enumerate}
    \item Mutation frequency (the fraction of sequenced individuals with the mutation), 
    \item Fraction of unique haplotypes in which the mutation occurs, 
    \item Number of countries in which the mutation occurs (in at least two sequences), and 
    \item Epi Score, an exponentially weighted mean of the above three epidemiological variables.
\end{enumerate} The Epi Score performed better than any other measure at predicting spread, with an AUROC of 0.99 in the RBD.

In particular, let $freq_i,hap_i,$ and $count_i$ denote the first three epidemiological variables, respectively, for a mutation $i.$ To find the Epi Score of $i,$ we first calculate $f_i, h_i,$ and $c_i,$ the percentiles for each value $freq_i,hap_i,$ and $count_i,$ respectively (compared to all mutations $i$). The three values $f_i,h_i,$ and $c_i$ are between 0 and 1 (0th to 100th percentile) for all $i.$ 

This can be achieved by sorting the mutations in increasing order by frequency ($freq$), haplotype occurrence ($hap$), and country occurrence ($count$), assigning the 0th percentile to the lowest ranks and 100th percentile to the highest. Then the Epi Score for the $i$th mutation is given by  
\begin{equation}\label{episcore}
    \textrm{Epi Score}_i=\frac{10^{f_i}+10^{h_i}+10^{c_i}}{3},
\end{equation} 
which gives a number between 1 and 10, with 1 signifying minimal risk and 10 presenting highest risk for spread.

By raising each percentile score as an exponential power, we allow for increased differentiation between high rankings (mutations with the most spread). For example, mutations in the 90th and 100th percentiles for mutation frequency would be assigned weights $10^{0.9}=7.9$ and $10^1=10,$ whereas mutations in the 40th and 50th percentiles would be assigned more similar weights of $10^{0.4}=2.5$ and $10^{0.5}=3.2.$

\section{Theory and Model}\label{sec:theory}
We introduce the Geo Score, modeled off the Epi Score (\ref{episcore}), as a method for disease risk assignment for different geographical regions rather than different mutations.

Instead of mutation frequency, haplotype occurrence, and country occurrence, we consider the three variables of vaccination rates $(vacc_i)$, population density $(dens_i)$, and socioeconomic status (quantified by median household income) $(ses_i),$ across regions within a city (in this case, we used the Census Bureau's ZIP Code Tabulation Areas). Again taking percentiles, this time with the lowest percentiles being assigned to the highest vaccinations and household incomes (and lowest population density), we calculate $v_i, d_i,$ and $s_i$, respectively. Then, for region $i,$ we test the performance of seven possible definitions of the Geo Score, each an exponential mean of different combinations of vaccination, population density, and socioeconomic status: 

\begin{align*}
    (\textrm{Geo Score 1})_i&=10^{v_i},\\
    (\textrm{Geo Score 2})_i&=10^{d_i},\\
    (\textrm{Geo Score 3})_i&=10^{s_i},\\
    (\textrm{Geo Score 4})_i&=\frac{10^{v_i}+10^{d_i}}{2},\\
    (\textrm{Geo Score 5})_i&=\frac{10^{v_i}+10^{s_i}}{2},\\
    (\textrm{Geo Score 6})_i&=\frac{10^{d_i}+10^{s_i}}{2},\\
    (\textrm{Geo Score 7})_i&=\frac{10^{v_i}+10^{d_i}+10^{s_i}}{3}.
\end{align*} 

To assess the predictiveness of these metrics, we consolidated ZIP code-level data on vaccination, population density, and household income in New York City from 2020, as well as up-to-date total positive test rates and death rates \cite{NYCCov23,ACSInc20,ACSPop20}. To ensure the same scale of comparison between the Geo Scores and outcomes (positivity and death rates), we also made two maps of outcome ``scores'' similar to Geo Scores 1, 2, and 3, raising 10 to the power of the test positive percentile and death percentile. Finally, we compared the resulting heatmaps of Geo Scores with the exponential positive test rate and death rate scores, and computed the mean errors by ZIP code between each score and the known results. 

\section{Results}\label{sec:results}
\begin{figure}[ht]
    \centering
    \includegraphics[width=10cm]{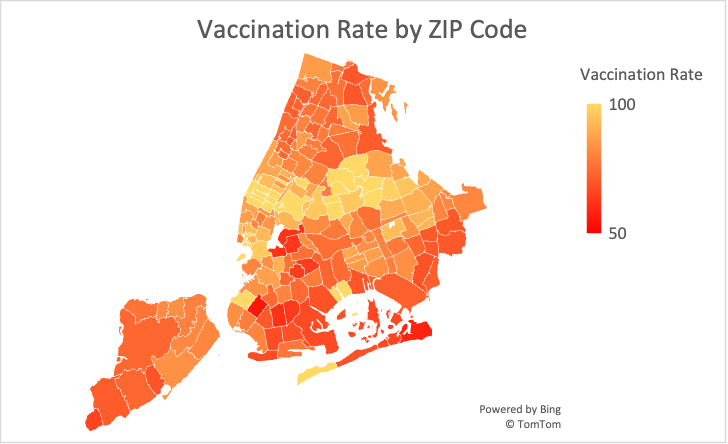}
    \caption{Vaccination rates in the ZIP codes of New York City, with the lowest rates presenting the highest risk (in red).}
    \label{fig:vaccination}
\end{figure}

\begin{figure}[ht]
    \centering
    \includegraphics[width=10cm]{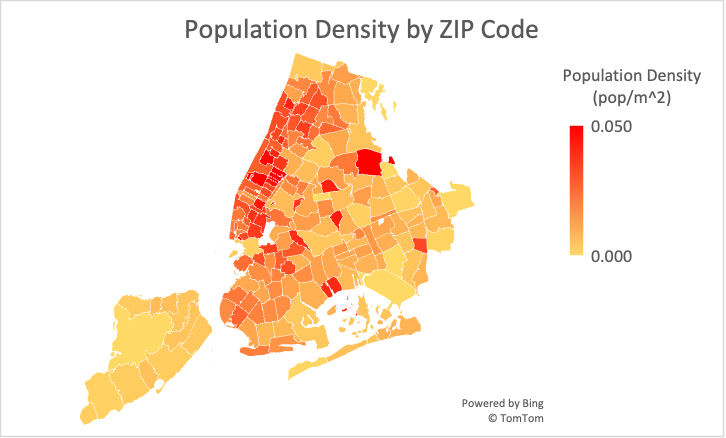}
    \caption{Population density in the ZIP codes of New York City, with the highest densities presenting the highest risk of transmission.}
    \label{fig:density}
\end{figure}

\begin{figure}
    \centering
    \includegraphics[width=10cm]{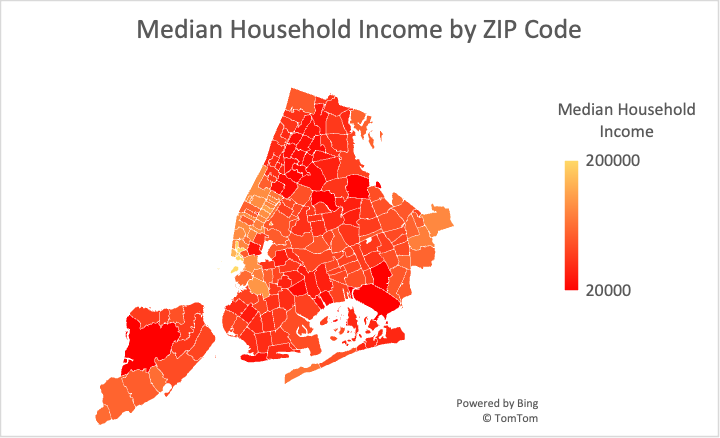}
    \caption{Median annual household income in New York City, with the lowest incomes reflecting poor living conditions and access to healthcare.}
    \label{fig:income}
\end{figure}

\begin{figure}
    \centering
    \includegraphics[width=10cm]{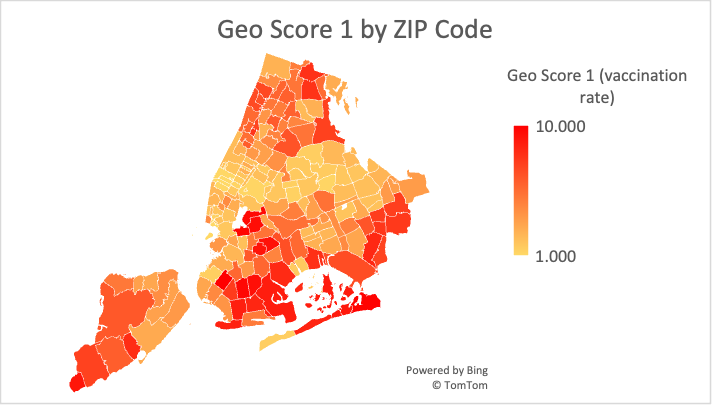}
    \caption{Geo Score 1 is an exponential of vaccination rates, which provides more differentiation between ZIP codes. A score of 10 poses the highest risk.}
    \label{fig:gs1}
\end{figure}

\begin{figure}
    \centering
    \includegraphics[width=10cm]{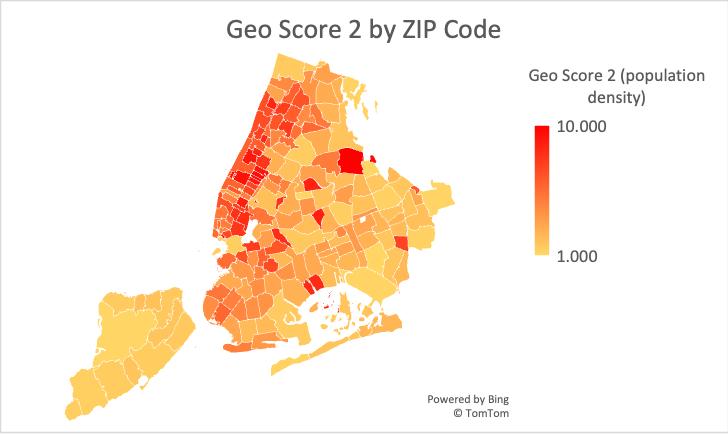}
    \caption{Geo Score 2 is an exponential of population density, and a score 10 represents the highest risk.}
    \label{fig:gs2}
\end{figure}

\begin{figure}
    \centering
    \includegraphics[width=10cm]{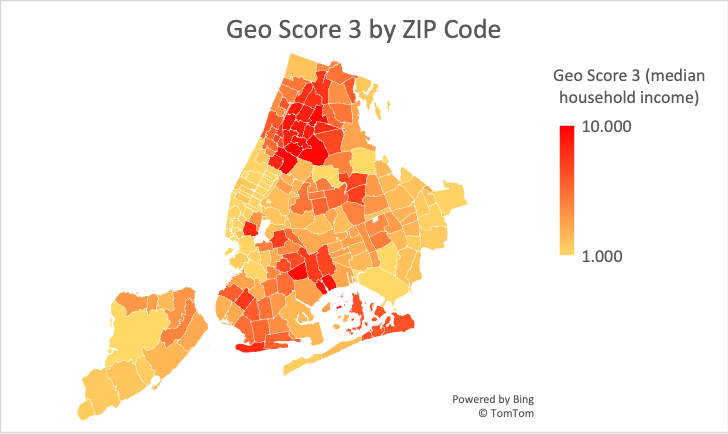}
    \caption{Geo Score 3, an exponential of median household income in the ZIP codes.}
    \label{fig:gs3}
\end{figure}

\begin{figure}
    \centering
    \includegraphics[width=10cm]{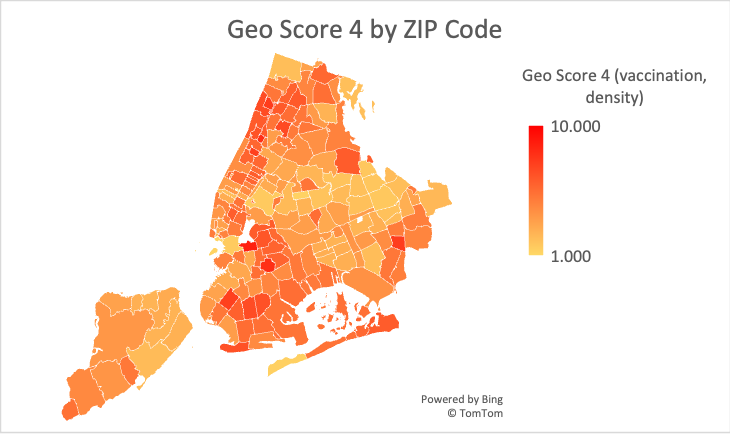}
    \caption{Geo Score 4 combines vaccination rates and population density in an exponentially-weighted mean.}
    \label{fig:gs4}
\end{figure}

\begin{figure}
    \centering
    \includegraphics[width=10cm]{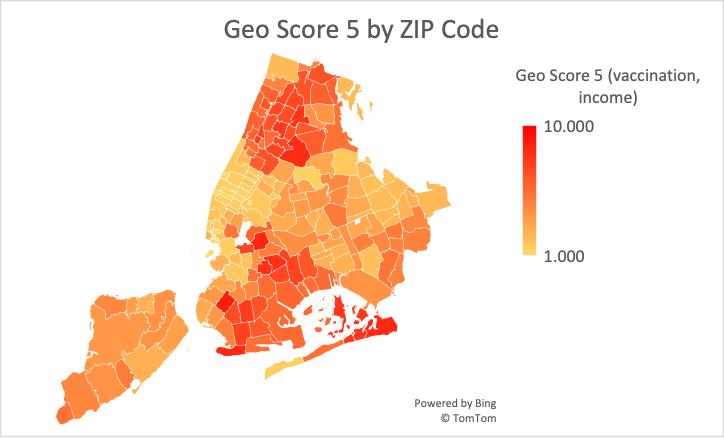}
    \caption{Geo Score 5 combines vaccination rates and household income in an exponentially-weighted mean.}
    \label{fig:gs5}
\end{figure}

\begin{figure}
    \centering
    \includegraphics[width=10cm]{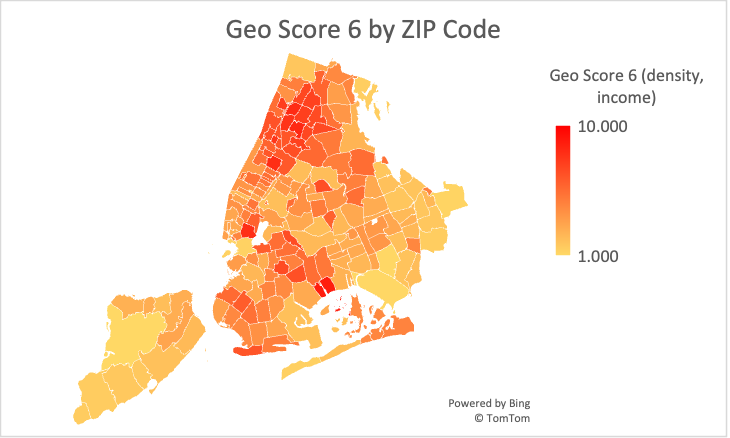}
    \caption{Geo Score 6 combines population density and household income in an exponentially-weighted mean.}
    \label{fig:gs6}
\end{figure}

\begin{figure}
    \centering
    \includegraphics[width=10cm]{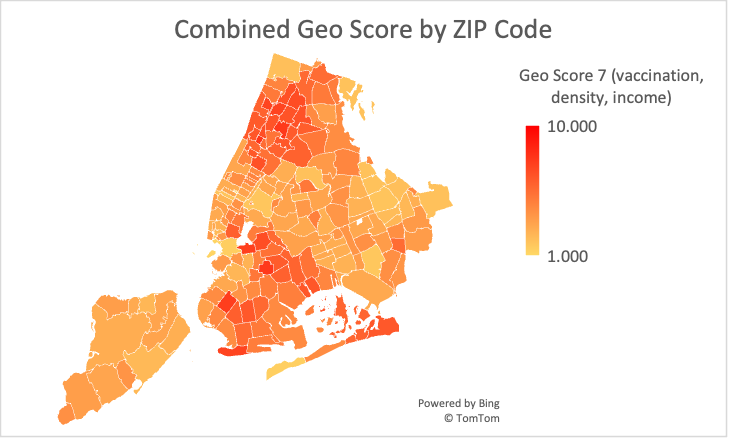}
    \caption{The combined Geo Score (Geo Score 7) exponentially averages all three variables: vaccination, population density, and household income.}
    \label{fig:gs7}
\end{figure}

\begin{figure}
    \centering
    \includegraphics[width=10cm]{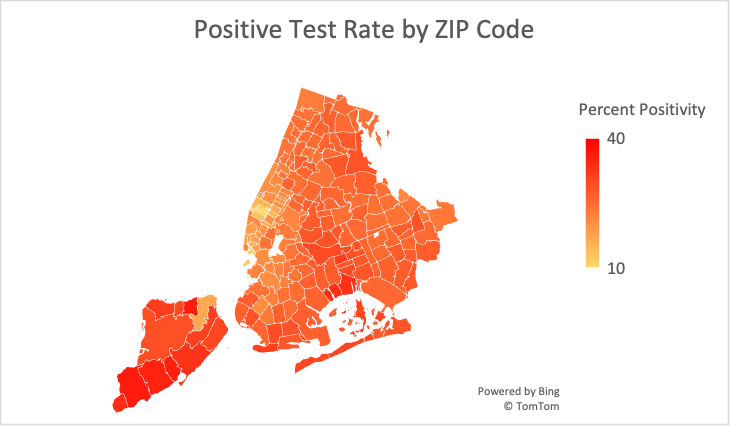}
    \caption{The positive rate of COVID-19 tests taken in the ZIP codes of New York City.}
    \label{fig:positive}
\end{figure}

\begin{figure}
    \centering
    \includegraphics[width=10cm]{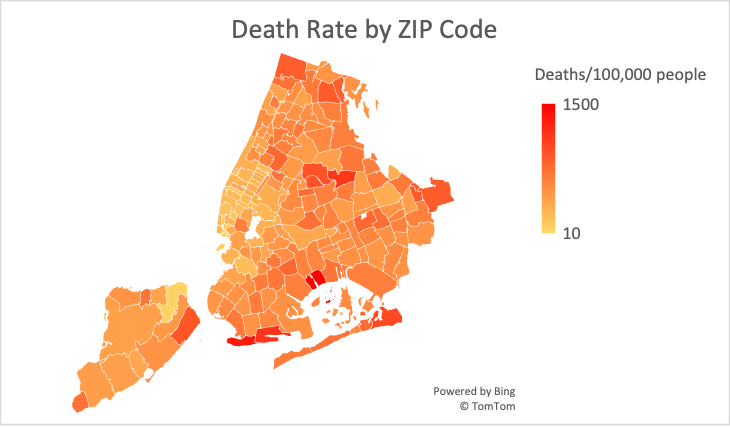}
    \caption{The death rate per 100,000 people in ZIP codes across New York City.}
    \label{fig:death}
\end{figure}

\begin{figure}
    \centering
    \includegraphics[width=10cm]{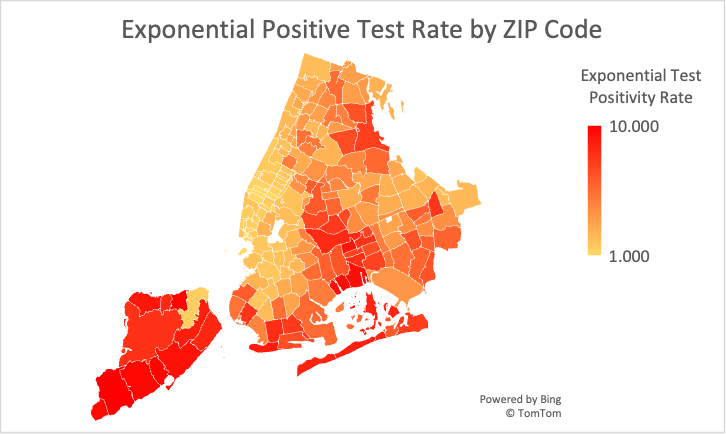}
    \caption{The exponential of ZIP codes' test positive rate percentile (modeled off the Epi Score).}
    \label{fig:exppositivity}
\end{figure}

\begin{figure}
    \centering
    \includegraphics[width=10cm]{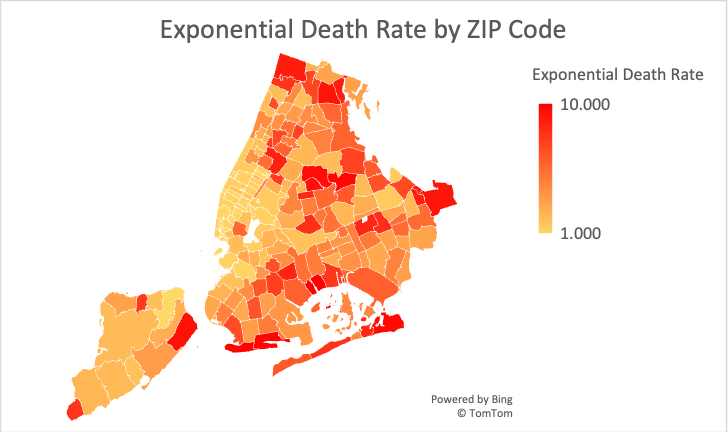}
    \caption{The exponential of New York City ZIP codes' death rate percentile.}
    \label{fig:expdeath}
\end{figure}

Below, we present our results in the form of 14 heatmaps (Figures \ref{fig:vaccination} through \ref{fig:expdeath}) showing vaccination rates, population density, median household income, and the seven Geo Scores, compared to positive test rates and death rates (in both their original and exponential forms), across the 214 ZIP Code Tabulation Areas of New York City. 

\begin{figure}
    \centering
    \includegraphics[width=7cm]{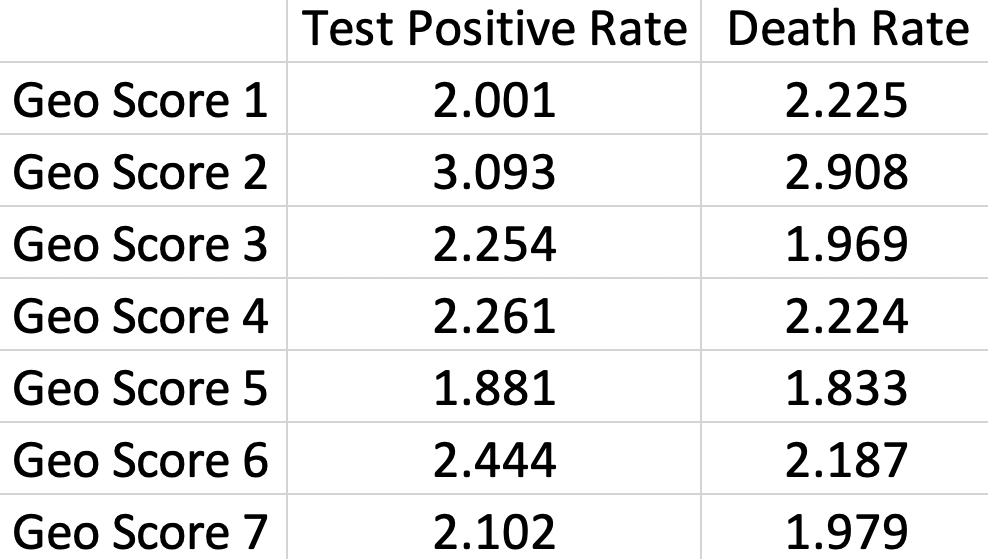}
    \caption{The average absolute difference across the ZIP codes between Geo Score and exponential outcomes, for each of the seven Geo Scores and two outcomes (positive test rate and death rate).}
    \label{fig:averageerror}
\end{figure}

\begin{figure}
    \centering
    \includegraphics[width=7cm]{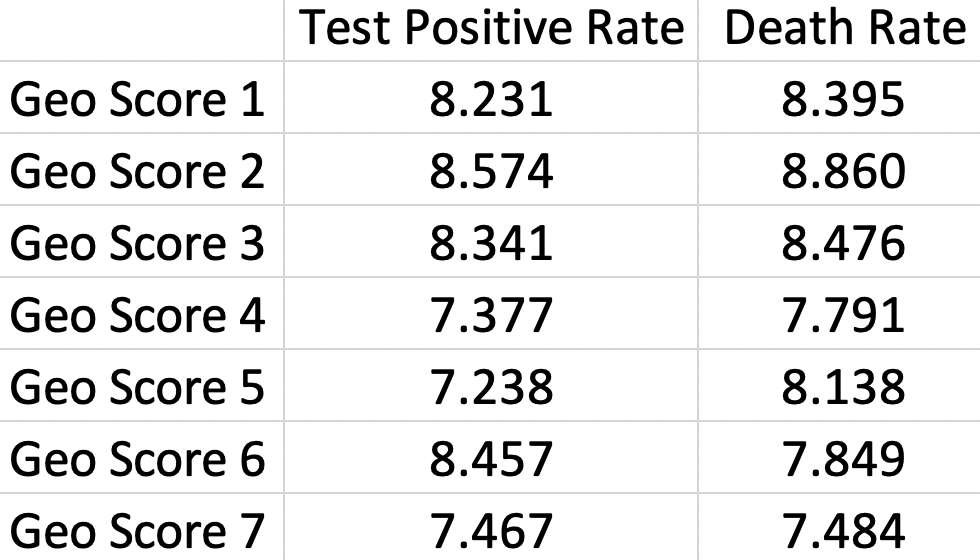}
    \caption{The maximum absolute difference between a Geo Score and exponential outcome value, for each of the seven Geo Scores and two outcomes.}
    \label{fig:maxerror}
\end{figure}

Figures \ref{fig:averageerror} and \ref{fig:maxerror} provide quantitative measures of the 7 Geo Scores' performance in predicting positive test rates and death rates, by taking the maximum and average of the absolute difference between the Geo Scores and outcome scores across the ZIP codes.


As expected, comparing Figures \ref{fig:vaccination} and  \ref{fig:gs1}, \ref{fig:density} and \ref{fig:gs2}, and \ref{fig:income} and \ref{fig:gs3} shows that applying the Epi Score model to the Geo Score allows for more differentiation between the ZIP codes. 

Additionally, in Figures \ref{fig:density} and \ref{fig:gs2}, we find that population density (both normal and exponential) is a poor indicator of positive test rates and death rates in the ZIP codes, ranking midtown Manhattan as one of the highest-risk areas in New York City. However, this region is designated as low-risk both by vaccination rate and median household income.

Thus, Geo Scores 4 and 6 (Figures \ref{fig:gs4} and \ref{fig:gs6}), which include population density in their average, also do not perfom well in predicting positivity and death rates. 

On the other hand, vaccination rates and median household income seem to predict test positivity relatively well, but death rates are more homogeneous between the ZIP codes, with only a few bright-red regions. We note that there were some gaps in the data for median household income from the five-year American Community Survey, specifically in ZIP Code Tabulation Area 10311, likely because of small land area and low population. Thus, in Figure \ref{fig:gs3}, the central area of Staten Island is inaccurately indicated with a Geo Score of 1.

Nevertheless, Geo Scores 5 and 7 (Figures \ref{fig:gs5} and \ref{fig:gs7}) are more predictive of positive rates and death rates (Figures \ref{fig:exppositivity} and \ref{fig:expdeath}) than the other combination Geo Scores, even including the data gaps.

\ref{fig:averageerror} and \ref{fig:maxerror} confirm our above qualitative analysis, showing that population density performs the worst, while vaccination and income are good indicators of spread. 

We can also see that, out of the single-variable Geo Scores, Geo Score 1 (representing vaccination rate) is the best predictor of the test positive rate, and Geo Score 3 (median household income) is the best predictor for death rate. This reflects the fact that vaccination is one of the most important factors in preventing infection, and that a higher socioeconomic status provides more resources to treat the disease and prevent death after infection. 

Combining Geo Scores 1 and 3 (vaccination and income), therefore, yields Geo Score 5, which is the best predictor overall, as measured both by average and maximum errors. Geo Score 5 also improves significantly upon every single-variable score.

Therefore, we find that the 2020 values of vaccination rates and median household incomes, as well as Geo Scores 5 and 7, perform the best in predicting accumulated positive test rates and death rates in the ZIP codes of New York City. 

\section{Tunable Weights}\label{sec:weights}
The Geo Scores above all assign equal weights to the variables included (whether there are one, two, or three). In this section, we consider varying the weights of variables to find a new, more accurate score for predicting test positive and death rates.

Let $p_1, p_2, p_3$ be the three distributions for vaccination rate, population density, and SES, respectively, and let $\alpha, \beta, \gamma \in [0,1]$ be the corresponding parameters such that $\alpha + \beta + \gamma = 1$. Then given these three empirical distributions, we look for the parameters such that the distribution 
$$
p=\alpha p_1 + \beta p_2 + \gamma p_3
$$
is as close as possible to the test positive rate distribution or the death rate distribution in the sense of minimizing the total $L_1$ distance (the sum of the absolute differences between the predicted and true outcome values across the ZIP codes in the distribution. This is equivalent to minimizing the mean absolute error (MAE).

\begin{figure}
    \centering
    \includegraphics[width=14cm]{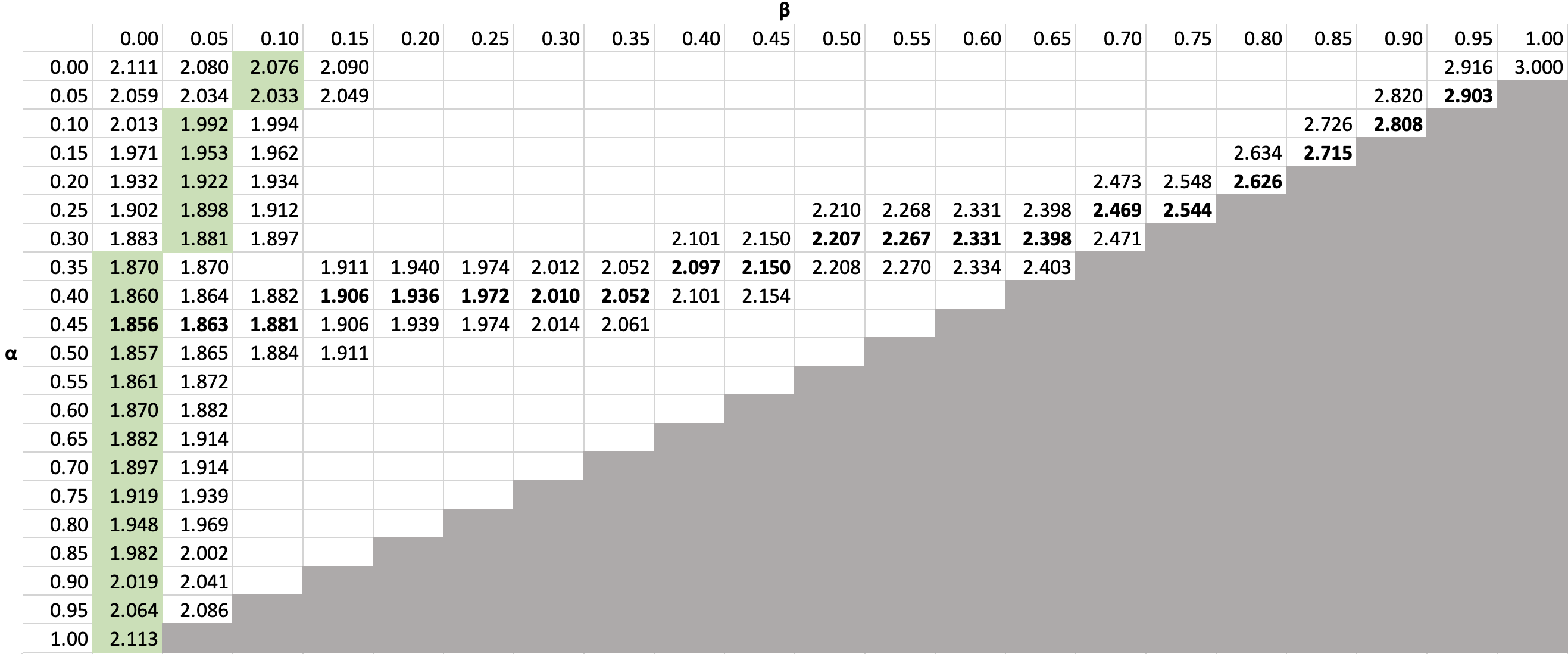}
    \caption{The average of the mean absolute errors when $p=\alpha p_1 + \beta p_2 + \gamma p_3$ is compared to the test positive rate distribution and the death rate distribution, for certain values of $\alpha$ and $\beta.$ The cells $(\alpha,\beta)$ that contain row/column minima, as well as their two surrounding cells, are filled in, to show the decrease and increase in error. Infeasible values of $\alpha$ and $\beta$ are greyed out to preserve $\gamma>0.$}
    \label{fig:avetable}
\end{figure}

We do this by calculating the average of the MAE of $\alpha p_1 + \beta p_2 + \gamma p_3$ when compared to the test positive rate distribution and the death rate distribution. In Figure \ref{fig:avetable}, we show the calculated average of the two mean errors, for $\alpha$ and $\beta$ going from 0 to 1 in increments of 0.05.

We find that in each row and each column (meaning, as we fix one parameter and vary the other), the average error decreases to a minimum, then increases again. 

We can prove this for the $L_2$ error. This makes sense because $$\alpha p_1+\beta p_2+\gamma p_3 = p_3+\alpha(p_1-p_3)+\beta(p_2-p_3),$$ and fixing $\alpha,$ this becomes $$c+\beta(p_2-p3),$$ where $c$ is the constant $p_3+\alpha(p_1-p_3).$ Thus, as $b$ varies, we are simply traveling along a line in 214-dimensional space (where 214 is the number of ZIP codes, $p_1,p_2,$ and $p_3$ are 214-dimensional vectors). Therefore, the $(L_2)$ distance from the true outcome distribution vector decreases to a minimum then increases (with the minimum when $\beta$ is such that we are at the foot of the perpendicular from the true outcome to the line). 

We can generalize to having $n$ weights $w_i$, where $n-1$ of them are free and the last is determined. Fixing each combination of $n-2$ of the $n-1$ free weights, we vary the last free weight to obtain a line. Then the $L_2$ distance to the true outcome vector decreases, then increases, as desired.

In Figure~\ref{fig:avetable}, the minimum errors are highlighted in green in each row, and bolded in each column. Thus, the intersection of the green and bolded trajectories is the absolute minimum error, at $\alpha=0.45,\beta=0,\gamma=0.55.$ 

Note, however, that this is an approximation, because we do not have an explicit continuous function for the average of the two mean absolute errors. Nevertheless, we notice that no row minimum has $\beta>0.1,$ indicating that $p_2$ (population density) is a poor predictor of test positive and death rates. Additionally, our absolute minimum has $\beta=0,$ and is very close to Geo Score 5 (which has $\alpha=0.5, \beta=0, \gamma=0.5$). As explored in the next section, however, optimizing for the MAE according to the test positive rate and the death rate separately requires unequal weights. 

\section{Gradient Descent}\label{sec:graddesc}
To find the values $\alpha$ and $\beta$ that minimize the $L_1$ error, we can use gradient descent. We let $R\left(\alpha,\beta\right)$ be the $L_1$ error function and $\alpha_n$ and $\beta_n$ be the values of the coefficients at ``time'' $t=n$ (at the $n$th iteration). Then, if the time  step is $\Delta t,$ we continuously update the coefficient values according to
\begin{align*}
\alpha_n&=\alpha_{n-1}-\left(\pdv{R}{\alpha}\right)_{\left(\alpha_{n-1}, \beta_{n-1}\right)}\cdot \Delta t,\\
\beta_n&=\beta_{n-1}-\left(\pdv{R}{\beta}\right)_{\left(\alpha_{n-1}, \beta_{n-1}\right)}\cdot \Delta t
\end{align*}
until the values converge (within a very small error), or the coefficients reach a boundary conditions ($\alpha,\beta,$ or $\gamma=1-\alpha-\beta$ decreases to 0 or increases to 1). Note that $R_\alpha$ and $R_\beta$ represent the partial derivatives of $R$ with respect to $\alpha$ and $\beta,$ respectively.

Given starting values $\alpha_0$ and $\beta_0,$ the only variables we need to evaluate to carry out the algorithm are $\pdv{R}{\alpha}$ and $\pdv{R}{\beta}.$ Let $R_i$ be the contribution to the total $L_1$ distance $R$ from each ZIP code $i$ in the training data. For the test positive rate, we have $$R(\alpha_n, \beta_n)=\sum_{i=1}^{k} R_i(\alpha_n,\beta_n)=\sum_{i=1}^k\big\lvert pos_i-\alpha(vacc_i)-\beta(dens_i)-(1-\alpha-\beta)(ses_i)\big\rvert,$$ where $i=1$ to $k$ represent the $k$ ZIP code areas in the training half of the data, and $pos_i, vacc_i, dens_i,$ and $ses_i$ denote the test positive rate, vaccination rate, population density, and median household income (SES) in area $i.$

Therefore, for each $i$ from 1 to $k,$ we have 
\begin{align*}
    \pdv{R_i}{\alpha}&=ses_i-vacc_i,\\
    \pdv{R_i}{\beta}&=ses_i-dens_i
\end{align*}
if $pos_i-\alpha(vacc_i)-\beta(dens_i)-(1-\alpha-\beta)(ses_i)>0,$ and 
\begin{align*}
    \pdv{R_i}{\alpha}(\alpha,\beta)&=vacc_i-ses_i,\\
    \pdv{R_i}{\beta}(\alpha,\beta)&=dens_i-ses_i
\end{align*} otherwise.

Summing the contributions across $i=1$ to $k$ for $\alpha=\alpha_n$ and $\beta=\beta_n,$ we can evaluate $$\left(\pdv{R}{\alpha}\right)_{\left(\alpha_{n}, \beta_{n}\right)}=\sum_{i=1}^k \left(\pdv{R_i}{\alpha}\right)_{\left(\alpha_{n}, \beta_{n}\right)},$$ and similarly $$\left(\pdv{R}{\beta}\right)_{\left(\alpha_{n}, \beta_{n}\right)}=\sum_{i=1}^k\left(\pdv{R_i}{\beta}\right)_{\left(\alpha_{n}, \beta_{n}\right)},$$ as desired.

To evaluate the performance of this gradient descent algorithm (implemented in Python), we split the dataset in half by alternating in the list of ZIP codes (sorted from smallest to largest), therefore covering all regions of the city in each half. We designate one half for ``training,'' to obtain the coefficients $\alpha$ and $\beta,$ then test these coefficients on the second half of the data and calculate the mean absolute error, which is equal to $\frac{1}{k}$ of the $L_1$ error. Doing this for both the test positive rate and death rate, we can then compare the obtained error against the test positive and death rate results of a standard multiple linear regression algorithm.

\begin{figure}
    \centering
    \includegraphics[width=11cm]{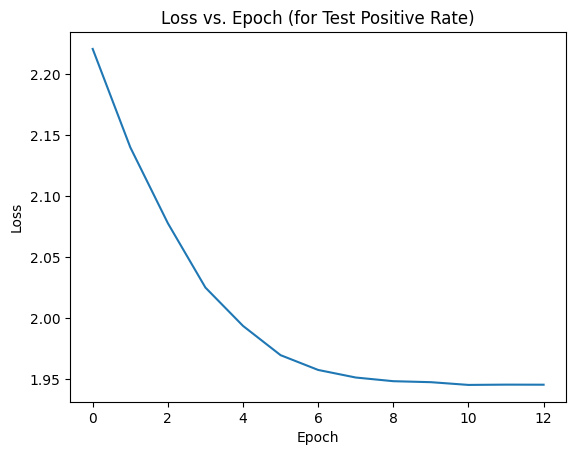}
    \caption{The MAE between the gradient descent's predicted test positive rate (according to $\alpha_n$ and $\beta_n$) and the true test positive rate across the training ZIP codes, at time $n$.}
    \label{fig:poserror}
\end{figure}

\begin{figure}
    \centering
    \includegraphics[width=11cm]{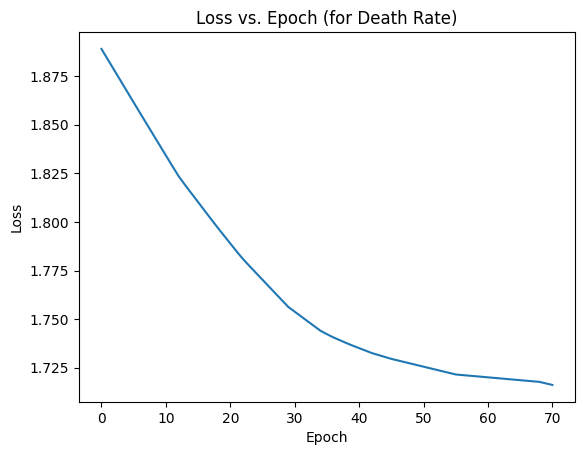}
    \caption{The MAE between the predicted death rate (according to $\alpha_n$ and $\beta_n$) and the true death rate across the training ZIP codes, at time $n$.}
    \label{fig:deatherror}
\end{figure}

We note that for different starting coefficients $(\alpha_0,\beta_0),$ we may obtain different ending (minimum $L_1$ error) coefficients; although $\beta_0\approx0$ in all cases, $\alpha_0$ varies between 0.3 and 0.7 when predicting the test positive coefficients, and between 0.6 and 0.8 for death rates. We find that both test positive and death rates do not depend much on population density, but that death rates are highly dependent upon socioeconomic status.

However, running the gradient descent algorithms for test positive and death rates, we always get minimum errors of roughly 1.94 and 1.71, respectively (out of 10). Compared to the linear regression model's mean absolute errors of 1.87 and 1.75, the gradient descent algorithm performs very well.

\section{Discussion}\label{sec:discussion}
In this paper, we utilize the idea of assigning scores to entities (in this case, geographical regions) to represent risk of infectious disease spread. By combining multiple scores of the same form, we create comprehensive scores to predict test positive rates and death rates (the outcomes indicative of disease spread), based on several significant variables: vaccination rate, population density, and socioeconomic status (quantified by median household income in the area). Although weighting these three scores equally yields decent results, solving for more appropriate weight coefficients allows for improved accuracy in predicting test positive and death rates.

We implement a gradient descent algorithm to solve for better-fitting weights, and compare the results to the linear regression model. However, the linear regression model of the test positive/death rate includes a constant term in addition to a weighted combination of vaccination, population density, and household income factors, and the coefficients for the three factors do not add up to 1. This causes the results of the linear regression to have little interpretability. In fact, the coefficient for the population density is predicted to be $-0.24<0$, meaning the lower the population density, the lower its population density risk score (Geo Score 2), but the higher the risk. This is likely an indication that, in New York City particularly, population density is positively correlated with other factors that lower risk when increased (like socioeconomic status or vaccination rate).

Therefore, the gradient descent algorithm performs almost just as well, if not better (in the case of death rates) than the linear regression model, and provides the added benefit of interpretability, while being easily generalizable to higher dimensions.

In this study, we focus on New York City and the spread of COVID from 2020-2021. However, our methods can be applied to other cities and diseases, which may uncover different weight coefficients indicating the varying significance of variables in other cities and regions. Further research can also be done to extract a minimal set of constraints given the spatiotemporal distribution of disease spread and death rates. An inverse problem of the traditional maximum entropy principle could allow the derivation of the minimum number of input variables, as well as their relative importance, needed to specify the resulting distributions of positive test rates and death rates.

\textbf{Acknowledgments.} C.L. would like to thank  Dr. Khovanova for her advice, and MIT PRIMES for the opportunity to pursue research.

\nocite{*}
\bibliographystyle{abbrv}
\bibliography{main}

\end{document}